\documentclass[amssymb, preprintnumbers, showpacs, showkeys,aps,prl,superscriptaddress,twocolumn,nobibnotes]{revtex4-1}

\usepackage{color} 
\usepackage{xcolor} 
\usepackage{hyperref}
\usepackage{slashed}
\usepackage{graphicx}
\usepackage{amsmath}
\usepackage{latexsym}
\usepackage{epstopdf}
\usepackage{amsmath}
\usepackage{enumitem}
\usepackage{amssymb,amsmath}
\usepackage{multirow}
\usepackage{mathtools}
\usepackage{bm}
\usepackage{amsfonts}
\usepackage{amssymb}
\usepackage{calligra}
\usepackage{calrsfs}
\usepackage{makecell}
\usepackage[normalem]{ulem}
\usepackage[USenglish,american]{babel}

\expandafter\ifx\csname package@font\endcsname\relax\else
 \expandafter\expandafter
 \expandafter\usepackage
 \expandafter\expandafter
 \expandafter{\csname package@font\endcsname}%
\fi
\begin{document}

\title{Self-induced Bose glass phase in quantum cluster quasicrystals}

\author{M. Grossklags}%
\email{matheus.grossklags@posgrad.ufsc.br}
\affiliation{Departamento de F\'\i sica, Universidade Federal de Santa Catarina, 88040-900 Florian\'opolis, Brazil}%

\author{M. Ciardi}%
\email{matteo.ciardi@unifi.it}
\affiliation{Dipartimento di Fisica e Astronomia, Universit\`a di Firenze, I-50019, Sesto Fiorentino (FI), Italy}%
\affiliation{INFN, Sezione di Firenze, I-50019, Sesto Fiorentino (FI), Italy}

\author{V. Zampronio}%
\email{vzampronio@fisica.ufrn.br}
\affiliation{Departamento  de  F\'isica  Te\'orica  e  Experimental,  Universidade  Federal  do  Rio  Grande  do  Norte}%

\author{F. Cinti}%
\email{fabio.cinti@unifi.it}
\affiliation{Dipartimento di Fisica e Astronomia, Universit\`a di Firenze, I-50019, Sesto Fiorentino (FI), Italy}
\affiliation{INFN, Sezione di Firenze, I-50019, Sesto Fiorentino (FI), Italy}
\affiliation{Department of Physics, University of Johannesburg, P.O. Box 524, Auckland Park 2006, South Africa}

\author{A. Mendoza-Coto}%
\email{alejandro.mendoza@ufsc.br}
\affiliation{Departamento de F\'\i sica, Universidade Federal de Santa Catarina, 88040-900 Florian\'opolis, Brazil}%
\affiliation{Max Planck Institute for the Physics of Complex Systems, Nothnitzerstr. 38, 01187 Dresden, Germany}

\begin{abstract}
    We study the emergence of Bose glass phases in self sustained bosonic quasicrystals induced by a pair interaction between particles of Lifshitz-Petrich type. By using a mean field variational method designed in momentum space as well as Gross-Pitaevskii simulations we determine the phase diagram of the model. The study of the local and global superfluid fraction allows the identification of supersolid, super quasicrystal, Bose glass and insulating phases. The Bose glass phase emerges as a quasicrystal phase in which the global superfluidity is essentially zero, while the local superfluidity remains finite in certain ring structures of the quasicrystalline pattern. Furthermore, we perform continuous space Path Integral Monte Carlo simulations for a case in which the interaction between particles stabilizes a quasicrystal phase. Our results show that as the strength of the interaction between particles is increased the system undergoes a sequence of states consistent with the super quasicrystal, Bose glass, and quasicrystal insulator thermodynamic phases.

\end{abstract}

\maketitle
\textit{Introduction} -- Physical systems forming self assembled patterns are ubiquitous in nature \cite{SeAn1995,AnRo2009}. These patterns are usually produced by an effective interaction between the many constituents of an ensemble presenting some sort of competition \cite{MeStNi2015,MeNiDi2019,Li2001,ZhMoJu2021}. The plethora of available frustrated interactions that naturally occur, or that can even be artificially engineered, combined with features like temperature, density of particles, defects or disorder, are responsible for the distinctive pattern formation observed in systems like  supersolids \cite{AnLi1969,Ch1970,Kim04correct,CiJaBoMiZoPu2010,BoPr2012,CiMaLePuPo2014}, quantum cluster crystals \cite{Ci2019,PuZiCi2020}, vortex lattices \cite{MaChWoDa2000,TsKaMa2002,AboShaeer2001,HeCiJaPuPo2012}, Bose glasses \cite{FiWeGrFi1989,KrTrCe1991,ScBaZi1991,DaZaSaZoLe2003,PhysRevLett.98.130404}, cavity QED mediated solids \cite{MiPiDoRi2021,KaPi2022,VaGuKrBaKoKeLe2018,GoLeGo2009,GoLeGo2010,ZhWaPo2021}, among many others \cite{GrMaEsHaBl2002,MeSt2012,BaMeSt2013,MeBaSt2017,MeBaNiDi2020}.

Generally, the most common scenario is the stabilization of some sort of periodic pattern at low temperatures while homogeneous states prevail in the high-temperature regime. Additionally, in the presence of a strong disorder or high concentration of system defects, even the low-temperature phase becomes disordered \cite{FiWeGrFi1989,FaLyGua2007}. In this context, an exotic realization of pattern formation lying between an ordered and a disordered phase is the so-called quasicrystals \cite{ShBlGrCa1984,LeSt1986}. These are ordered patterns that do not break translational symmetry but display a long-range orientational order whose symmetry is forbidden to any other periodic crystal in that specific dimension \cite{Se1996}. Such properties are the ultimate result of building a pattern with a wave vector basis whose dimension is higher than the dimension of the system. 

In the context of soft matter physics, extensive work has been done on characterizing quasicrystal structures formed by many particles, namely, cluster quasicrystals \cite{LiDi2007,Do2011,BaDiLi2011}. In a foundational work by Barkan et al. \cite{BaEnLi2014}, it was shown that cluster quasicrystals can be stabilized by an effective pair potential containing two properly selected competing length scales in the presence of moderate thermal fluctuations. In the quantum case, a recent work \cite{CoTuZaMeMaCi2022} has shown that to obtain a dodecagonal quasicrystal density pattern in a Bose-Einstein condensate, analog to the classical case but stabilized by quantum fluctuations, three local minima must be correctly positioned in the Fourier transform of the effective pair interaction potential. To the best of our knowledge, potentials with two competing length scales could only stabilize quasicrystals in the regime of moderate to low intensity of quantum fluctuations \footnote{We have considered LPG interaction potentials with two minima and performed an analysis using the variational mean field method and GPE simulations and verified that the cases in which the dodecagonal quasicrystal corresponds to the ground state of the system are located at moderate or high values of $\lambda^{2}\rho U$. We have observed that for some regions in this parameter space, the GPE simulations show that for a dodecagonal quasicrystal initial condition, the system evolves preserving the pattern but its stationary state doesn't correspond to the lowest energy configuration. This implies meta stability of the structure in the aforementioned cases. In this sense, the definitive criterion to determine the ground state is the value of the energy per particle of the system.}.

More recently, a number of publications\cite{YaKhBr2019,YaGiSa2020,GaYaSa2021} motivated by experimental feasibility have considered the stabilization of quasicrystal phases by the application of an external quasiperiodic potential in systems in which particles repel each other by contact interactions.  One of the main results of these studies is the prediction of a Bose glass (BG) phase in such models. The BG phase is a well-established phase for models of interacting bosons in the presence of some sort of disorder, it usually occurs as an intermediate phase between the Mott insulator state and the superfluid phase \cite{FiWeGrFi1989}. In terms of its physical properties, it is a compressible insulator that hosts localized superfluid regions \cite{SvBaPr2015}. Due to the fact that the loss of coherence leading to the insulating behavior is a consequence of the lack of periodicity in the system, it is expected for the energy spectrum to manifest some sort of fractal structure, instead of energy bands as in the case of systems subjected to periodic external potentials \cite{FaLyGua2007,YaKhBr2019,YaGiSa2020}. In the described scenario, an intriguing possibility is the existence of a BG phase for quantum self-assembled quasicrystals in Bose-Einstein condensates at zero temperature.

In terms of experimental perspectives, it is important to mention that although we consider a specific mathematical form of the pair interaction potential, the main ingredient causing the physical behavior reported is the existence of a properly constructed three minima structure in its Fourier transform. The achievement of sign-changing interaction potentials in real space is the fundamental feature to generate the necessary frustration to stabilize quantum cluster quasicrystals and other exotic modulated states. In a number of recent works, effective interactions of this type mediated by photons in cavity QED systems have been used to produce different modulated patterns in ultra-cold quantum gases  \cite{MiPiDoRi2021,KaPi2022,VaGuKrBaKoKeLe2018,GoLeGo2009,GoLeGo2010,ZhWaPo2021}.

In this letter, we study the emergence of a self-organized BG phase in a Bose-Einstein quasicrystal condensate stabilized by the simultaneous role of quantum delocalization and a two-body interaction with multiple characteristic length scales. The BG phase is identified by studying simultaneously the local and global superfluidity since in our case the system is intrinsically compressible in all regimes of parameters. The ground state phase diagram of the system is studied using two complementary techniques, a variational approach \cite{CoTuZaMeMaCi2022} devised in momentum space, free of the effects of finite system sizes as well as periodic boundary conditions and the more standard Gross-Pitaevskii equation (GPE) simulations \cite{Pi1961,Gr1963}, as a secondary test for the precision of the main analytical method. Additionally, we perform continuous space Path Integral Monte Carlo (PIMC) simulations \cite{Ce1995,BoPrSv2006b} for a finite portion of the system subjected to a confining potential in order to mimic possible experiments. The computational results for the finite system confirm qualitatively the main findings of the analytical study for the system in the thermodynamic limit.

\begin{figure}
    \centering
    \includegraphics[width=0.70\textwidth]{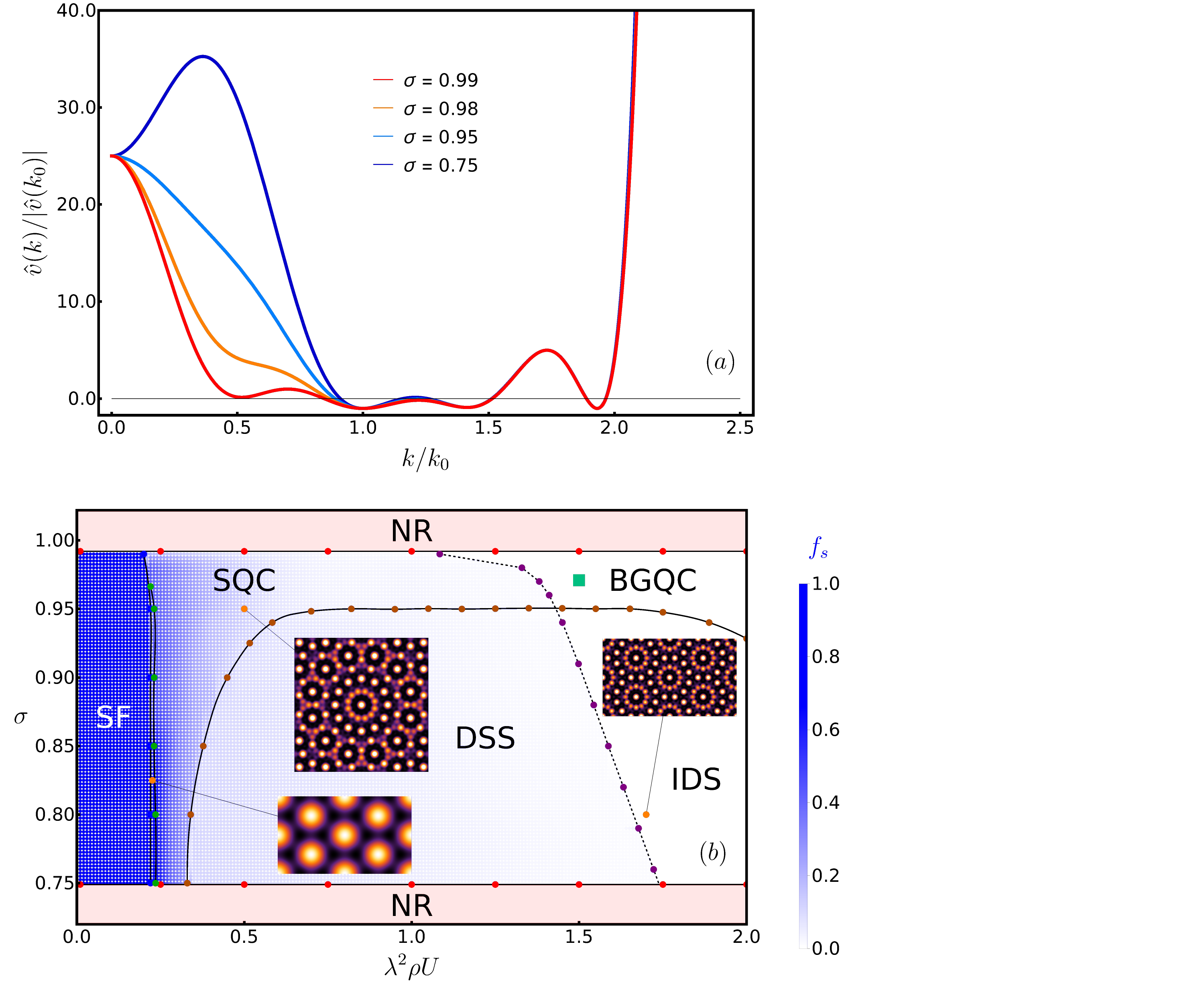}
    \caption{\textbf{(a)} Pair interaction potential $\hat{v}(k)$ for four different values of the parameter $\sigma$ considered in the phase diagram of our model. \textbf{(b)} Ground state phase diagram in the $\sigma$ versus $\lambda^{2}\rho U$ plane. The small blue points with varying opacity represent the value (see legend) of the superfluid fraction at each point in the diagram. The non-relevant area (NR) corresponds to regions in which our LPG model is inconsistent with the constraints imposed to the potential. The local properties of the green square are explored in Figure \ref{fig:image2}(a)-(c).}
    \label{fig:image1}
\end{figure}

\textit{Model and approach} -- We consider a two-dimensional system of $N$ bosonic atoms of mass $m$ at zero temperature. The particles interact through an isotropic Lifshitz-Petrich-Gaussian \cite{BaEnLi2014} (LPG) pair interaction potential defined in momentum space as 
\begin{equation}
    \hat{v}\left(k\right)=Ve^{-k^{2}\sigma^{2}}\sum\limits_{n=0}^{8}d_{n}k^{2n}.
\label{model}
\end{equation}
This kind of pair interaction potential was originally proposed due to mathematical convenience to study models that stabilize periodic and aperiodic patterns in soft matter systems \cite{BaEnLi2014}. We set the coefficients $d_n$ in our model such that $\hat{v}(k)$ has a three minima structure. Two of these minima, at $k_0$ and $\sqrt{2+\sqrt{3}}k_0$, are degenerate, and a third one at $\sqrt{2}k_0$, is included to enhance the stability of the dodecagonal quasicrystal phase presented by this model in certain regions of the parameter space \cite{CoTuZaMeMaCi2022}. In Figure~\ref{fig:image1}(a), we show the family of potentials considered in the present work, the numerical expression of $\hat{v}(k)$ is obtained fixing the values of this function at $\{0,k_0,\sqrt{2}k_0,\sqrt{3}k_0,\sqrt{3+\sqrt{2}}k_0\}$ as well as the parameter $\sigma$.

In the limit of weak interactions, after expressing the spatial coordinates and the energy of the system in units of $\lambda=2\pi/k_0$ and $\epsilon=\hbar^2/m\lambda^2$, respectively, we can write the energy per particle of the condensate as 
\begin{equation}
    \begin{split}
        \frac{E}{N}=&\ \frac{1}{2}\int\frac{d^{2}\bm{x}}{A}\lvert\bm{\nabla}\psi\left(\bm{x}\right)\rvert^{2}+\frac{\lambda^{2}\rho U}{2}\int\frac{d^{2}\bm{x}d^{2}\bm{x}'}{A}v\left(\bm{x}-\bm{x}'\right)\\
        &\times\lvert\psi\left(\bm{x}\right)\rvert^{2}\lvert\psi\left(\bm{x}'\right)\rvert^{2},
    \end{split}
\end{equation}
where $A$ stands for the area of the system and $\psi(\bm{x})$ represents the normalized condensate wave function satisfying $\int dx\lvert\psi\left(\bm{x}\right)\rvert^2=A$. In these conditions, the local density of particles is given by $\rho(\bm{x})=\rho\lvert\psi(\bm{x})\rvert^{2}$, where $\rho$ corresponds to the dimensionless average density of particles. Additionally, the dimensionless parameter $U=V/\epsilon$ sizes the intensity of the pair interaction potential.

\begin{figure*}
    \centering
     \includegraphics[width=1.00\textwidth]{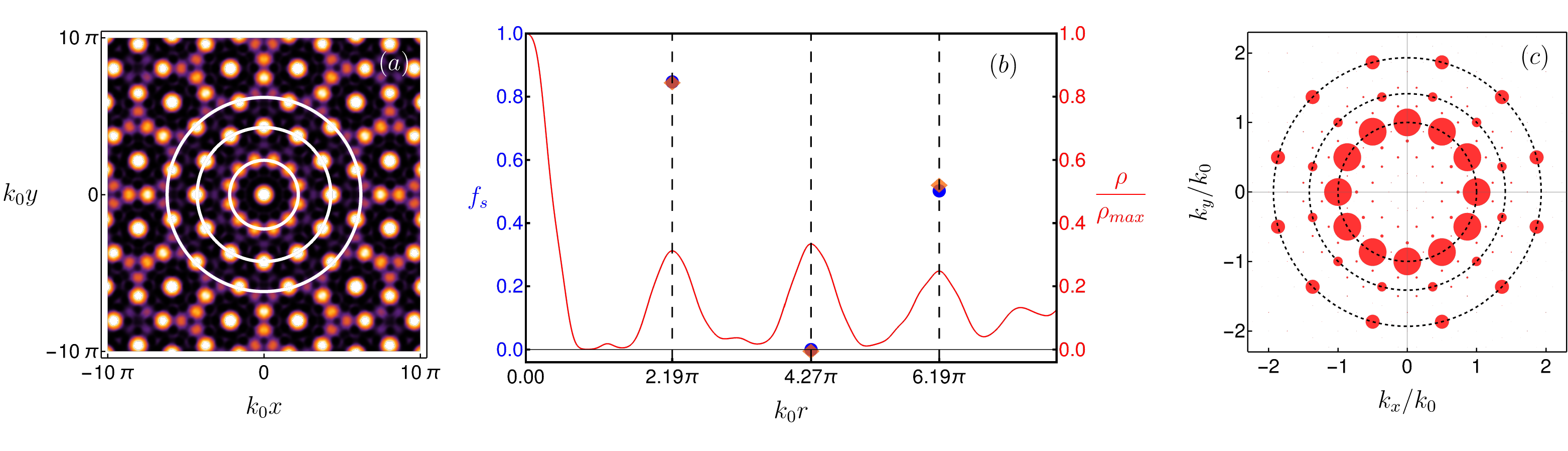}
    \caption{\textbf{(a)} Real space density plot of the quasicrystalline structure in the insulating phase for the interaction potential parameter $\sigma=0.97$ and density $\lambda^{2}\rho U=1.50$ obtained from the variational approach. The three white rings represent the circles of interest at which we evaluate the local superfluid fraction of the system in (b). \textbf{(b)} Local superfluid fraction obtained for the white circles indicated in (a) using the variational approach (blue circles) and GPE simulations (orange diamonds). The red line indicates the average radial density. \textbf{(c)} Structure factor of the quasicrystalline structure at the same parameter values indicated in (a). Here the three black dashed circles represent the three minima structure of the LPG interaction potential at $\{k_{0},\sqrt{2}k_{0},\sqrt{3+\sqrt{2}}k_{0}\}$.}
    \label{fig:image2}
\end{figure*}
  
To investigate the ground state of the system, we minimize the energy per particle functional considering two complementary approaches. The first one uses a many-mode Fourier expansion for the wave function of the form \cite{ZhMaPo2019,CoTuZaMeMaCi2022} 
\begin{equation}\label{ground_state_wave_function}
    \psi\left(\boldsymbol{x}\right)=\frac{c_{0}+\frac{1}{2}\sum\limits_{j\neq0}^{j_{\text{max}}}c_{j}\cos{\left(\boldsymbol{k}_{j}\cdot\boldsymbol{x}\right)}}{\left(c_{0}^{2}+\frac{1}{4}\sum\limits_{j\neq0}^{j_{\text{max}}}c_{j}^{2}\right)^{1/2}},
\end{equation}
to determine the ground state configuration. The set of Fourier amplitudes $\{c_j\}$ and the characteristic wave vectors of the pattern defined by $\{\bm{k}_j\}$ are taken as variational parameters for the minimization process \cite{PrSeBr2018,PrSeBr2018b}. The limit of the sum $j_{\text{max}}$ in  Eq.\ref{ground_state_wave_function} is only formal. The Fourier expansion for the ground state wave function in Eq.\ref{ground_state_wave_function} considers a set of vectors $\{\textbf{k}_j\}$ that is two dimensional and is constructed recursively, using $n_{c}$ successive combinations of the wave vectors basis of the pattern. All periodic solutions were constructed using a $n_{c}=30$, while the quasiperiodic solution uses $n_{c}=8$. Such truncation criterion showed good convergence properties in all tests performed. 
We consider a group of possible ans\"atze for the ground state wave function consistent with all modulated patterns reported in the literature for two and three-length scale potentials with the same properties considered here \cite{LiPe1997,BaEnLi2014}. One of the advantages of this expansion is that it allows us to obtain an exact analytical expression for the functional energy per particle in terms of the Fourier amplitudes and wave vectors.

To complement this approach and to check that the set of possible solutions considered in our variational calculations is complete, we also performed GPE simulations evolving in imaginary time \cite{DeHoJo2013}. We probed different initial conditions in which we seed a portion of a given structure considered in our variational approach and confirm the meta stability or not of a certain type of solution checking if the initial state evolves into the full structure. Furthermore, we also considered initial conditions with and without rotational symmetry and observed that in all cases the evolution of the system lead us to a state in which the local texture is consistent with the kind of modulated pattern we considered in our variational approach. In order to characterize the ground state of the system with high accuracy, we considered a substantial Fourier basis to guarantee energy convergence for the expansion of all possible solutions. These convergence tests were performed at specific $\lambda^{2}\rho U$ values along the whole phase diagram presented. We observed that this method produces lower energy values than those resulting from evolving the GPE in imaginary time, in most cases by a small relative error of $10^{-2}$.

Additionally, we complement our study of the modulated phases by estimating the superfluid fraction ($f_{s}$) using the upper bound criterion provided by Legget \cite{Le1970}. However, due to mathematical convenience, we take advantage of a recent result \cite{CoTuZaMeMaCi2022} showing that the original expression proposed by Legget for periodic and quasiperiodic patterns is equivalent to
\begin{equation}\label{legget_criterion}
    f_{s}=\left(\int \frac{d^2\bm{x}}{A}\vert\psi\left(\bm{x}\right)\vert^{-2}\right)^{-1}.
\end{equation}
By including the characterization of the superfluid fraction in the phase diagram using GPE simulations, we can identify regions in which the modulated phases exist in a superfluid $\left(f_s > 10^{-2}\right)$ or insulating state $\left(f_s<10^{-2}\right)$. It is expected that a calculation beyond mean field for the insulating regime will produce a strictly zero value for the superfluid fraction since it is well established that mean-field calculations in these cases smooth out the superfluid-insulator phase transition \footnote{A comparison between the results from PIMC and GPE simulations for interacting quantum gases in quasiperiodic traps have shown that in the regime in which mean-field calculations produces a low superfluid fraction of the order $f_s\approx10^{-2}$, PIMC calculations exhibit a phase transition to a strictly globally insulating state.}. This implies that in the regions in which we have the coexistence of modulated phases and superfluidity, we have supersolid, or supersolid-like phases, while in the regions in which the modulated phases are concomitant with the absence of superfluidity, we have insulators, at least from the global perspective. However, as we will see shortly, the absence of global superfluidity does not rule out the existence of local superfluidity in certain ring structures of the modulated patterns.  

\textit{Phases of the model} -- The results of the variational mean field method in combination with GPE simulations are shown in the ground state phase diagram in Figure \ref{fig:image1}(b). Here we can observe six distinct phases: the homogeneous superfluid phase (SF), a supersolid hexagonal phase, which is only stable for low enough values of $\sigma$ in a quite narrow region of densities, a super dodecagonal quasicrystal (SQC), a supersolid decorated hexagonal phase (DSS), an insulating decorated solid (IDS) and a Bose glass quasicrystal (BGQC). Furthermore, we would like to add that we decided to evaluate the superfluid fraction for the phase diagram in Figure~\ref{fig:image1}(b) from GPE simulations since the variational mean field method produces abnormally low values of this quantity at large enough $\lambda^{2}\rho U \left(>1.0\right)$. This issue is related to the extremely low-density regions between the cluster of particles. As a consequence, a huge number of Fourier modes are needed in the expansion to accurately describe the ground state wave function in these depleted regions. While this behavior affects the superfluid fraction, it does not have a strong impact on the energy values of a given pattern and in this sense, the ground state phase diagram is completely reliable.

To better understand the nature of the insulating phases in our diagram, we now turn our attention to their local properties. In Figure~\ref{fig:image2}(a)-(c) we present a study of the ground state configuration at $\sigma=0.97$ and density $\lambda^{2}\rho U=1.50$, which corresponds to a value (green square in Figure \ref{fig:image1}(b) within the insulating BGQC. Figure \ref{fig:image2}(a) shows the central region of the density profile in this case. It is possible to observe how the quasicrystal structure is formed by a composition of different kinds of corona structures, formed by twelve clusters of particles distributed over a circumference. The central coronas of the pattern were highlighted using white rings with radius $k_{0}r=2.19\pi, 4.27\pi$, and $6.19\pi$, respectively. In Figure \ref{fig:image2}(b), we compute simultaneously the average density profile of the pattern in the radial direction and the normalized superfluid fraction corresponding to each of the highlighted central corona structures. The blue circles correspond to the superfluid fraction calculated using the density profile from the variational approach while the orange diamonds use the GPE simulation results. As we can observe both methods predict equivalent results for this quantity even in the insulating phases. This is a consequence that in regions with a significant density of particles the minimization variational method accurately describe the density profile since they contribute the most to the energy of the system. It is interesting to notice that while the global superfluid fraction obtained from both methods is effectively zero, we find a high value for the local superfluidity on the first and third corona structure, while the second one reproduces the behavior of the global superfluidity within the insulating regime. 

Although not shown, a similar analysis in the decorated phase for the local superfluidity produces alike results, indicating that this kind of property in our case should be mostly associated with the kind of pair interaction potential we are considering and not exclusively with a particular phase. Indeed, the pair potential in our case punishes with a high energy cost all Fourier modes of the density pattern with characteristic wave vector $k\geq2k_0$. In Figure \ref{fig:image2}(c), we show the Fourier modes distribution for the density pattern in Figure \ref{fig:image2}(a), where the radius of the circles are proportional to the absolute value of the corresponding Fourier amplitudes. As we can observe the main excited modes coincide with the minima of the interaction potential, signaled by the black dashed circumferences. Moreover, this also confirms that although present in our variational ans\"atze, modes with high momentum are in fact strongly suppressed, which should reflect in a large $\lambda^{2}\rho U$ extension of the regime hosting local but not global superfluid properties.

\begin{figure*}
\centering
\includegraphics[width=\textwidth]{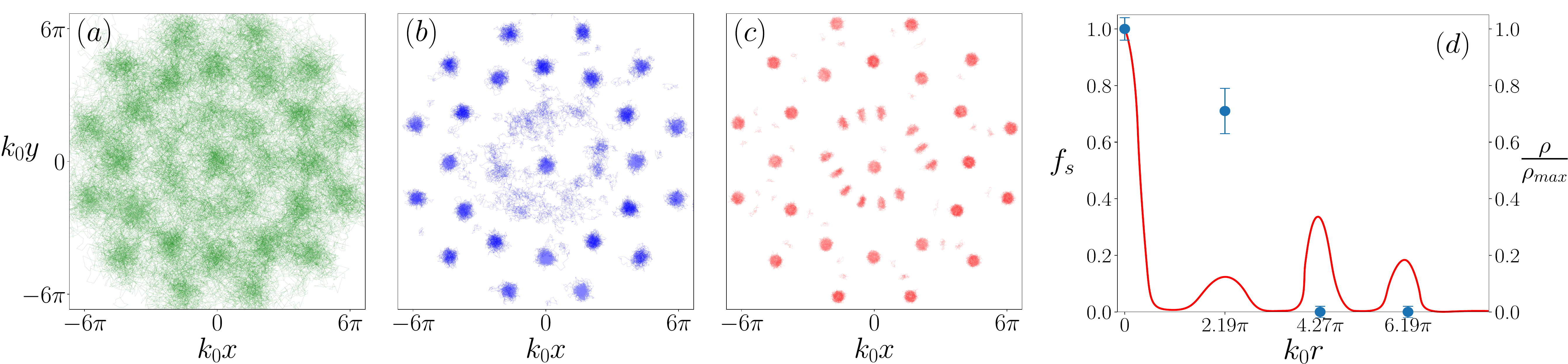}
    \caption{\textbf{(a-c)} PIMC world line configurations, for the \textbf{(a)} superfluid, \textbf{(b)} Bose glass, and \textbf{(c)} insulating phase. \textbf{(d)} Normalized radial density distribution in the Bose glass phase (red solid line) and zonal superfluidity in the coronas (blue circles).}
    \label{fig:image3}
\end{figure*}

The phenomenology described above, in addition to the well-established literature related to the Bose glass phase in disordered and quasiperiodic systems, allows us to infer that the global insulating BGQC observed is actually a BG phase \cite{SvBaPr2015}. Typically a Bose glass is distinguished from a Mott insulator by the presence of finite small compressibility, however, our model is by definition always compressible, and in this sense, the usual criterion to differentiate these phases in lattice models does not apply to our case. In addition to the local and global superfluidity behavior as criteria to determine whether we are in the presence of a BG phase, we should also consider the nature of the particle-hole excitation spectrum. In the case of the insulating BGQC, the lack of periodicity in the system will produce gapless particle-hole excitations between neighboring superfluid clusters. Despite the zero gap, excitations will only occur locally due to the lack of coherence over a large distance making it globally insulating as in a Bose glass. In the case of a decorated solid independent of its superfluid properties, it is well established that the periodicity of the structure will produce an energy spectrum with a band structure which is incompatible with the fractal energy spectrum of a Bose glass.

\textit{PIMC's results} -- We employ a continuous-space PIMC method to deeply investigate the proposed system by implementing the efficient worm algorithm \cite{BoPrSv2006b}. In this form PIMC provides an accurate description of a quantum system by estimating the key thermodynamic observables. As known, the methodology works in a regime of finite temperature albeit a proper extrapolation to the $T$=0 limit has been correctly obtained in numerous systems \cite{0034-4885-75-9-094501}. 
In the present work, we simulate a fixed number of quantum particles to obtain density and superfluid fraction. When addressing quasicrystalline geometries, periodic boundary conditions are not reliable and it is preferred to use external confinement \cite{PhysRevA.105.L011301,e24020265}. In this case, we employed an external potential of the form $V(r) \propto (r/r_c)^{\alpha}$ with $\alpha$ a large integer e.g. $\alpha = 30$, and $r_c$ large enough to contain the first four coronas; this approximates an infinite wall at $r_c$.

In order to confirm the stability of the structures seen in the mean-field methods, and to evidence deviations, we perform simulations in continuous space starting from a density profile derived from the mean-field ground state. The starting configuration is derived as follows. First, we select a portion of the ground state, as a circle of radius $R$ centered in the symmetry center of the quasicrystal. We pick a density $n$, and from it, we derive the particle number as $N = n / \pi R^2$. Inside the circle, we use a thresholding algorithm to determine regions of high density, corresponding to the clusters forming the quasicrystal structure. To each cluster $j$, we assign a weight $w_j$, proportional to the integrated density on the cluster; a number of particles $N_j$ are then assigned to each cluster proportionally to $w_j$, approximating them so that each cluster has an integer number of particles. Finally, the particles are positioned randomly within each cluster, proportionately to the density. The goal of this process is to ensure that different clusters have a balanced number of particles, without introducing strong fluctuations of the particle number in the clusters.

Aiming to verify the mean-field results about local superfluidity in the coronas, 
we measured the zonal superfluid fraction and compressibility for each of them, following the procedure outlined in Ref.~\cite{e24020265}. As shown in Fig.\ref{fig:image3}(a-c), we were able to distinguish a superfluid, insulating, and Bose glass phase, similar to what is shown in Ref.~\cite{PhysRevA.105.L011301}, where a Bose glass is denoted by localized particles and local superfluity only in certain kind of coronas of the QC structure. For a choice of parameters, we observe a transition of the same kind as seen in mean field, except that the transition from insulator to Bose glass is shifted to smaller values of $\lambda^{2}\rho U$, e.g. $\lambda^{2}\rho U\approx 0.02$ for $\sigma = 0.97$. An example of the behavior of the local superfluidity in the different coronas withing the Bose glass phase is shown in Fig\ref{fig:image3}(d). As we can observe the local superfluid properties obtained in simulations are qualitatively the same to those reported in the mean-field study. The insulating behavior displayed by the last corona of the simulated system is already expected due the absence of the subsequent quasicrystal structure in the simulations. The discrepancy between mean-field and PIMC for the boundary of the Bose glass phase appears to be due to the presence of the confining external potential, which is extremely different from the effective potential created by the rest of the quasicrystal structure. We expect that as we enlarge the system--thus bringing it closer to the infinite-size homogeneous system--the transition should recover the values seen in the mean-field approach.

\textit{Conclusions and discussion} -- In the present work, we studied the ground state properties of a two-dimensional bosonic gas interacting via a LPG pair potential. The selected model presents three characteristic length scales properly chosen in order to favor the stabilization of a dodecagonal quasicrystal structure. The phase diagram of the system is investigated using two mean-field complementary methods. The first one follows a direct variational approach to determine the optimal modulated ground state, given as an expansion in Fourier modes for each kind of modulated solution. To test and complement these results we perform extensive Gross-Pitaevskii simulations along the phase diagram. While the first variational method was employed to construct the modulations phase diagram, Gross-Pitaevskii simulations were used to obtain an accurate description of the superfluid properties deep into the modulated region of the phase diagram. Additionally, we performed PIMC simulations of a restricted region of the quasicrystal structure to verify the presence of a Bose glass phase as well as the stability of the self-induced quasicrystal structure.

Our results shows that for high enough $\lambda^{2}\rho U$ values the self-induced quasicrystal phase eventually lose its global superfluid properties. However, in this regime it is able to host local superfluidity in certain rings structures of the quasiperiodic pattern. This kind of behavior is taken as a signature of a Bose glass phase, considering that in the present system, due to absence of a rigid lattice, compressibility will be always finite. The main ingredient for the phenomenology reported is the presence of a pair interaction with the adequate three minima structure in momentum space, which is currently within experimental reach \cite{MiPiDoRi2021,KaPi2022,VaGuKrBaKoKeLe2018,GoLeGo2009,GoLeGo2010,ZhWaPo2021}. Additionally, PIMC simulations confirmed qualitatively the scenario obtained from the mean-field study. Within both methods, the first corona of the quasicrystalline pattern is able to retain superfluidity while the second corona of the structure mirrors the global insulating behavior. Finally, we believe that the results presented in this work motivates new experiments in the direction of producing a self-induced quasicrystal phase. Not only due to the exotic nature of such a phase but also due to the possibility of an eventual production of the elusive Bose glass phase.

\providecommand{\noopsort}[1]{}\providecommand{\singleletter}[1]{#1}%

\end{document}